%% file: LDPCCC.tex
\documentclass[conference]{IEEEtran}

\usepackage[cmex10]{amsmath}
\usepackage{amssymb}
\usepackage{amsthm}
\usepackage{array}
\usepackage[english]{babel}  
\usepackage{bbm}
\usepackage[font=footnotesize]{caption}
\usepackage{cite}
\usepackage{dsfont}
\usepackage{enumerate}
\usepackage{epsfig} 
\usepackage{epstopdf}
\usepackage{fancyhdr}
\usepackage{fixltx2e}
\usepackage[T1]{fontenc}
\usepackage{graphicx}
\usepackage[latin1]{inputenc}                       
\usepackage{mathtools}
\usepackage{latexsym}
\usepackage{pgfplots}
\usepackage{syntonly}
\usepackage{url}
\usepackage{tikz}
\usetikzlibrary{arrows.meta}
\usetikzlibrary{shapes,arrows}
\usepackage{verbatim}
\usepackage{wasysym}
\usepackage{xcolor}
\usepackage{relsize}
\newtheoremstyle{custom}
{} 				% Space above
{} 				% Space below
{} 				% Body font
{} 				% Indent amount
{\bfseries} 		% Theorem head font
{.} 				% Punctuation after theorem head
{.25em} 			% Space after theorem head
{} 				% Theorem head spec (can be left empty, meaning `normal')
\theoremstyle{custom}

\newtheorem{lemma}{Lemma}

\newtheorem{definition}{Definition}

\newtheorem*{theorem*}{Theorem}
\newtheorem*{lemma*}{Lemma}
\newtheorem*{proposition*}{Proposition}
\newtheorem*{definition*}{Definition}
\newtheorem*{example*}{Example}
\newtheorem*{remark*}{Remark}
\newtheorem*{corollary*}{Corollary}

\newcommand{\bcode}{{\mathcal B}}
\newcommand{\ccode}{{\mathcal C}}

\newcommand{\y}{\mathbf{y}}

\newcommand{\h}{\mathbf{h}}

\newcommand{\HP}{\mathbf{H}}

\makeatletter
\let\l@ENGLISH\l@english
\makeatother

\title{Time-invariant LDPC convolutional codes}
\author{\IEEEauthorblockN{Dimitris Achlioptas\IEEEauthorrefmark{1}, Hamed Hassani\IEEEauthorrefmark{2}, Wei Liu\IEEEauthorrefmark{3}, and R\"{u}diger Urbanke\IEEEauthorrefmark{3}}
\IEEEauthorblockA{\IEEEauthorrefmark{1}Department of Computer Science, UC Santa Cruz, USA\\ Email: \texttt{achlioptas@cs.ucsc.edu}}
\IEEEauthorblockA{\IEEEauthorrefmark{2}Department of Computer Science, ETH Z\"{u}rich, Switzerland\\ Email: \texttt{hamed@inf.ethz.ch}}
\IEEEauthorblockA{\IEEEauthorrefmark{3}School of Computer and Communication Sciences, EPFL, Switzerland\\ Emails: \texttt{\{wei.liu,ruediger.urbanke\}@epfl.ch}} }

\begin{document}
\maketitle

%%%%%%%%%%%%%%%%%%%%%%%%%%%%%%%%%%%%%%%%%%%%%%%%%%%%%%%%%%%%%%%%%%%%%%%%%%
%--------------------------------Abstract--------------------------------%
%%%%%%%%%%%%%%%%%%%%%%%%%%%%%%%%%%%%%%%%%%%%%%%%%%%%%%%%%%%%%%%%%%%%%%%%%%

\begin{abstract}
Spatially coupled codes have been shown to universally achieve the
capacity for a large class of channels. Many variants of such codes
have been introduced to date.  We discuss a further such variant
that is particularly simple and is determined by a very small number
of parameters. More precisely, we consider {\em time-invariant}
low-density convolutional codes with very large constraint lengths.

We show via simulations that, despite their extreme simplicity,
such codes still show the {\em threshold saturation} behavior known
from the spatially coupled codes discussed in the literature.
Further, we show how the size of the typical minimum stopping set
is related to basic parameters of the code. Due to their simplicity
and good performance, these codes might be attractive from an
implementation perspective.  \end{abstract}

%%%%%%%%%%%%%%%%%%%%%%%%%%%%%%%%%%%%%%%%%%%%%%%%%%%%%%%%%%%%%%%%%%%%%%%%%%
%-------------------------------Section++--------------------------------%
%%%%%%%%%%%%%%%%%%%%%%%%%%%%%%%%%%%%%%%%%%%%%%%%%%%%%%%%%%%%%%%%%%%%%%%%%%

\section{Introduction} 
\label{sec:intro}

\input{introduction}

%%%%%%%%%%%%%%%%%%%%%%%%%%%%%%%%%%%%%%%%%%%%%%%%%%%%%%%%%%%%%%%%%%%%%%%%%%
%-------------------------------Section++--------------------------------%
%%%%%%%%%%%%%%%%%%%%%%%%%%%%%%%%%%%%%%%%%%%%%%%%%%%%%%%%%%%%%%%%%%%%%%%%%%

\section{Binary Linar Codes} 
\label{sec:blc}

\input{blc.tex}

%%%%%%%%%%%%%%%%%%%%%%%%%%%%%%%%%%%%%%%%%%%%%%%%%%%%%%%%%%%%%%%%%%%%%%%%%%
%-------------------------------Section++--------------------------------%
%%%%%%%%%%%%%%%%%%%%%%%%%%%%%%%%%%%%%%%%%%%%%%%%%%%%%%%%%%%%%%%%%%%%%%%%%%
\section{Construction} 
\label{sec:codeconstruction} 
\label{sec:construction}

\input{construction}

%%%%%%%%%%%%%%%%%%%%%%%%%%%%%%%%%%%%%%%%%%%%%%%%%%%%%%%%%%%%%%%%%%%%%%%%%%
%-------------------------------Section++--------------------------------%
%%%%%%%%%%%%%%%%%%%%%%%%%%%%%%%%%%%%%%%%%%%%%%%%%%%%%%%%%%%%%%%%%%%%%%%%%%

\section{Stopping Sets} 
\label{sec:ss}

\input{stopping_sets}

%%%%%%%%%%%%%%%%%%%%%%%%%%%%%%%%%%%%%%%%%%%%%%%%%%%%%%%%%%%%%%%%%%%%%%%%%%
%-------------------------------Section++--------------------------------%
%%%%%%%%%%%%%%%%%%%%%%%%%%%%%%%%%%%%%%%%%%%%%%%%%%%%%%%%%%%%%%%%%%%%%%%%%%

\section{Simulation Results}
\label{sec:simulations}

\input{simulation}

%%%%%%%%%%%%%%%%%%%%%%%%%%%%%%%%%%%%%%%%%%%%%%%%%%%%%%%%%%%%%%%%%%%%%%%%%%
%-------------------------------Section++--------------------------------%
%%%%%%%%%%%%%%%%%%%%%%%%%%%%%%%%%%%%%%%%%%%%%%%%%%%%%%%%%%%%%%%%%%%%%%%%%%

\section{Conclusions}
\label{sec:concl}

We introduced {\em time-invariant} low-density convolutional codes.
These codes are defined by a very small number of bits. We have
seen that despite their simplicity these codes perform very well.
We have given some simple lower bound on the minimum stopping
set size of such codes which grows exponentially in the number
of shift-invariant parity-checks.

%%%%%%%%%%%%%%%%%%%%%%%%%%%%%%%%%%%%%%%%%%%%%%%%%%%%%%%%%%%%%%%%%%%%%%%%%%
%-------------------------------Section++--------------------------------%
%%%%%%%%%%%%%%%%%%%%%%%%%%%%%%%%%%%%%%%%%%%%%%%%%%%%%%%%%%%%%%%%%%%%%%%%%%

\section*{Acknowledgement}

The work of W.~Liu and R.~Urbanke is supported by grant No. 200021\_166106 
of the Swiss National Science Foundation. The work of H. Hassani is partially supported by a fellowship from Simons Institute for the Theory of Computing, UC Berkeley. 

%%%%%%%%%%%%%%%%%%%%%%%%%%%%%%%%%%%%%%%%%%%%%%%%%%%%%%%%%%%%%%%%%%%%%%%%%%
%------------------------------Bibliography------------------------------%
%%%%%%%%%%%%%%%%%%%%%%%%%%%%%%%%%%%%%%%%%%%%%%%%%%%%%%%%%%%%%%%%%%%%%%%%%%

\bibliographystyle{IEEEtran}
\bibliography{lth,lthpub,ieeexplore}

%%%%%%%%%%%%%%%%%%%%%%%%%%%%%%%%%%%%%%%%%%%%%%%%%%%%%%%%%%%%%%%%%%%%%%%%%%
%--------------------------------DOCUMENT--------------------------------%
%%%%%%%%%%%%%%%%%%%%%%%%%%%%%%%%%%%%%%%%%%%%%%%%%%%%%%%%%%%%%%%%%%%%%%%%%%
\end{document}

%% file: introduction.tex
%!TEX root = LDPCCC.tex

Spatially coupled codes have been shown to universally achieve the
capacity for a large class of channels, \cite{FZ99,LSZC10,5513587,6887298}.
Many variants of such codes have been introduced to date. For the
purpose of analysis it is convenient to consider highly random
ensembles. For implementation purposes it is convenient to eliminate
as much randomness as possible, e.g., by considering constructions
based on protographs.

We ask how much ``randomness'' is required for such codes to perform
well. A natural setting is to look at the origins of spatially
coupled codes and to consider {\em low-density parity-check} (LDPC) convolutional codes
with large constraint lengths. But rather than considering {\em time-variant} LDPC convolutional codes, 
we consider {\em time-invariant} such codes (and hence the number of parameters that describe such codes is very small). 
As we will see, even these extremely simple codes exhibit in simulations the
{\em threshold saturation} phenomenon that is well-known from the
standard spatially coupled codes discussed in the literature.

Let us recall. {\em LDPC block codes} are linear
codes defined by parity-check matrices where the number of non-zero
elements per parity-check is small and independent of the blocklength. 
Due to the small degrees of the checks such codes can be
decoded ``well'' by a message-passing decoder. To be more precise:
For well-designed ensembles, the threshold (when we consider ensembles
of codes whose blocklength tends to infinity) under belief propagation
decoding is close to the Shannon limit, \cite{LMSSS97,RSU01}.
Nevertheless this threshold is typically strictly smaller than the
threshold that would be achievable if we were able to implement
maximum {\em a posteriori} (MAP) decoding, which is the optimal decoding strategy.

{\em LDPC convolutional codes} can be seen as
convolutional codes (with very large constraint lengths) defined
by parity-check equations with only a small number of non-zero taps
(and the number of taps is independent of the constraint length).
Standard convolutional codes are typically decoded by means of the
Viterbi algorithm, whose complexity is exponential in the constraint
length. For the codes that we consider the constraint length is
hundreds or even thousands. Decoding via the Viterbi algorithm is
therefore not feasible. But, just as for LDPC block codes, 
these codes can be decoded ``well'' via a message-passing
algorithm due to the low-density nature of the parity-checks. There
is one big difference to block codes, however. Whereas for block
codes the iterative decoding threshold is generically strictly smaller than
the MAP threshold, for convolutional codes with a proper ``seeding''
at the boundary, the two thresholds coincide. This phenomenon has
been dubbed {\em threshold saturation} in the literature, \cite{5513587}
and has been observed (and in some cases proved) for various
spatially coupled ensembles.

For the LDPC convolutional codes that are discussed in the literature,
it is assumed that the filter coefficients are {\em time variant}.
For some instances (like the ensembles that are most suitable for
proofs) the amount of randomness that is required scales with the
length of the code. For other instances, in particular for the type
of spatially coupled codes that are defined by ``unwrapping'' a
block code, the randomness is proportional to the memory of the
code, \cite{Tan81}.

We consider {\em time-invariant} LDPC convolutional
codes. Each such code is defined by only $\Theta(n (n-k)\log_2(W))$ bits, 
where $W$ is a ``window'' design parameter that determines
the ``effective'' blocklength of the code. The parameter $n$ is
equal to the number of streams of the code, and $k/n$ is the rate
of the code. We show via simulations that, despite their extreme
simplicity, such codes show the {\em threshold saturation} behavior
known from standard spatially coupled codes discussed in the
literature. Further, we show how the typical minimum stopping set
size is related to the basic parameters of the code. Due to their
simplicity and good performance, these codes might be attractive
from an implementation perspective.

This paper has two objectives. First, we want to investigate how
little randomness is needed in order to construct good spatially
coupled codes. Second, we want to point out the strong analogy that exists between
block and convolutional codes. In both cases, there are no polynomial
time algorithms known that accomplish decoding close to capacity
when we consider the dense case. But when we restrict ourselves to
codes defined by sparse parity-check constraints then the message-passing
algorithm works well. The major difference is that for block codes
the iterative decoding threshold is generically strictly worse than the MAP
threshold whereas for convolutional codes they generically coincide.

In order to bring out this analogy even clearer we will start from
scratch and quickly review some basics of coding theory. This is
done in Section~\ref{sec:blc}. In Section~\ref{sec:construction}
we then describe the exact ensemble that we consider. It is
particularly simple and suitable for analysis. In Section~\ref{sec:ss}
we show that the size of the minimum stopping sets that we should
expect in such a code grows exponentially in the degree of the check
nodes. Finally, in Section~\ref{sec:simulations} we present some
basic simulations. We limit ourselves to the binary erasure channel (BEC)
for ease of exposition but the general phenomenon is not limited
to this channel.

%% file: blc.tex
%!TEX root = LDPCCC.tex

\subsection{Block Codes}

\begin{definition}[$(n, k)$ Block Code -- Parity View]
An $(n, k)$ linear binary block code $\bcode$ can be defined by
\begin{align*}
\bcode = \{\y \in \{0, 1\}^n: \HP \y^{\top} = 0 \},
\end{align*}
where
\begin{align*}
\HP & = 
\left(
\begin{array}{cccc}
h_{1,1} & h_{1,2}& \cdots & h_{1,n} \\
\vdots & \vdots & \ddots & \vdots \\
h_{n-k,1} & h_{n-k,2}& \cdots & h_{n-k,n} \\
\end{array}
\right)
\end{align*}
is a binary matrix of dimensions $(n-k) \times n$. It is called the {\em
parity-check} matrix.  
\end{definition}
%Note: If we choose the elements of $\HP$ independently from each
%other and uniformly at radom from $\{0, 1\}$ then the resulting
%ensemble has very similar properties than the previously discussed
%generator ensemble (even though it is not identical).

%The relationship between these two view points is well known. In
%particular, if we choose the generator matrix $\G$ to be systematic,
%i.e., to have the form $\G = (\I_{k \times k}, \PP_{k \times (n-k)})$,
%then the corresponding parity-check matrix $\HP$ has the form $\HP
%= (\PP_{k \times (n-k)}^{\top}, \I_{(n-k) \times (n-k)})$.

There are very good codes in such an ensemble. In fact, if we allow
MAP decoding then such codes achieve capacity for a large class of
channels (e.g., the class of binary-input memoryless output-symmetric
channels).

Unless such codes have further structure, no algorithms are known
that can accomplish decoding close to the threshold in polynomial
time. In fact, the best known generic decoding algorithms have
complexity $O(2^{\text{min}\{k, n-k\}})$.
% \footnote{Do you need a $O(\cdot)$ here?}
% We will get back to this point shortly.

\subsection{Convolutional Codes}
Let ${\mathcal F}(D)$ be the set of formal power sums in the
indeterminate $D$ with binary coefficients $\sum_{i \geq 0} a_i
D^i$ and only non-negative powers of $D$.

%\begin{definition}[$(n, k)$ Convolutional Code -- Generator View]
%An $(n, k)$ convolutional code $\ccode$ can be defined by
%\begin{align*}
%\ccode = \{\y(D) \in {\mathcal F}(D): \y(D) = \x(D) \G(D); \x(D) \in {\mathcal F}(D)\},
%\end{align*}
%where
%\begin{align*}
%\G(D) & = 
%\left(
%\begin{array}{cccc}
%\g_{1,1}(D) & \g_{1,2}(D)& \cdots & \g_{1,n}(D) \\
%\vdots & \vdots & \ddots & \vdots \\
%\g_{k,1}(D) & \g_{k,2}(D)& \cdots & \g_{k,n}(D) \\
%\end{array}
%\right)
%\end{align*}
%is a matrix of dimensions $k \times n$ with entries $\g_{i, j}(D)$
%that are polynomials in the indeterminate $D$. It is called the
%{\em generator} matrix.

%The {\em memory} $m$ of the code is defined as
%\begin{align*}
%m = \max_{1 \leq i \leq k; 1 \leq j \leq n} \text{deg}[\g_{i, j}(D)]
%\end{align*}
%and the constraint length $c$ is defined as
%\begin{align*}
%c = \sum_{i=1}^{k} \max_{1 \leq j \leq n} \text{deg}[\g_{i, j}(D)].
%\end{align*}
%\end{definition}
%Note that in the above definition we have defined the code by
%describing its {\em encoder}. This is the {\em generator}
%point of view and it mirrors the way we described a (linear) block
%code defined by a generator matrix.

%In practice we of course do not use infinite streams but we consider
%finite length portions with an appropriate termination.  We ignore
%this issue in our notation for the purpose of simplicity

%Rather than describing a code by its generator matrix we can describe
%it by its parity-check matrix.

\begin{definition}[$(n, k)$ Convolutional Code -- Parity View]
An $(n, k)$ convolutional code $\ccode$ can be defined by
\begin{align*}
\ccode = \{\y(D) \in {\mathcal F}(D): \HP(D) \y(D) = 0\},
\end{align*}
where
\begin{align*}
\HP(D) & = 
\left(
\begin{array}{cccc}
\h_{1,1}(D) & \h_{1,2}(D)& \cdots & \h_{1,n}(D) \\
\vdots & \vdots & \ddots & \vdots \\
\h_{n-k,1}(D) & \h_{n-k,2}(D)& \cdots & \h_{n-k,n}(D) \\
\end{array}
\right)
\end{align*}
is a matrix of dimensions $(n-k) \times n$ with entries $\h_{i,
j}(D)$ that are polynomials in $D$. It is called the {\em parity-check}
matrix.
\end{definition}
%Just as for block codes we have a simple relationship (which formally
%looks the same) between the generator matrix and the parity-check
%matrix if we start with a generator matrix that is in systematic
%form.

The {\em memory} $M$ of the code is defined as
\begin{align*}
M = \max_{1 \leq i \leq n-k; 1 \leq j \leq n} \text{deg}[\h_{i, j}(D)],
\end{align*}
and the {\em constraint length} $L_c$ is defined as
\begin{align*}
L_c = \sum_{i=1}^{n-k} \max_{1 \leq j \leq n} \text{deg}[\h_{i, j}(D)].
\end{align*}

Optimal decoding of such codes is typically accomplished by running
the so-called Viterbi algorithm. Its complexity is exponential in
the constraint length. 
% We will get back to this point shortly. 
Similarly to block codes, convolutional codes are capacity-achieving for a
wide range of channels under optimal decoding at least 
if we allow the filter tap coefficients to be time-variant, \cite{ViO79}.
% at least if we allow the polynomials to be functions of {\color{red} time \cite{ViO79}.} 
% \footnote{Remove the comma.
% What is ``time''? It is somehow odd to, on one hand, give such a very basic definition
% of convolutional codes, but on the other not explain what time is. Maybe its best to suppress
% the ``, at least if ...'' part.}

\subsection{Low-Density Parity-Check Block Codes} Much of the advance
in modern coding theory and practice has come about by looking at
{\em sparse} versions of block codes, \cite{Gal62,BGT93,RiU08}.
More precisely, we consider codes defined via parity-check matrices
where the parity-check matrix is {\em sparse} / has a {\em low-density} of non-zero entries. 
Many versions of such LDPC block codes have been discussed in the literature.
In the simplest case we can assume that every row of $\HP$ has $r$
non-zero entries and every column has $l$ non-zero entries. This
is called an $(l, r)$-regular LDPC code.

For LDPC block codes decoding is typically done via a message passing
algorithm. As we discussed in the introduction, such codes, if well
designed, have thresholds very close to the Shannon capacity.
But typically the threshold under iterative decoding is strictly
smaller than the threshold under MAP decoding, \cite{MMU04}. 

\subsection{Low-Density Parity-Check Convolutional Codes}

As for block codes, we can consider convolutional codes defined by 
low-density parity-check matrices. More precisely,
we let the memory tend to infinity but we keep the number of non-zero
tap coefficients per row constant, independent of the memory.

We then use a standard message-passing algorithm on
the Tanner graph of the code, rather than a Viterbi algorithm. Since
the check degrees are constant, the complexity of the message-passing
algorithm is linear in the overall length of the code.

It is important to point out that the codes we consider are similar
to the standard codes discussed in the literature, i.e., they are
{\em LDPC convolutional codes}. But we consider {\em time-invariant} codes 
whereas typically {\em time-variant} versions are discussed in the literature. 
The exact construction we consider is described in the next section.

%% file: construction.tex
%!TEX root = LDPCCC.tex

In the previous section we discussed already the generic class of
LDPC convolutional codes. The following construction is particularly
convenient from the point of view of analysis. But we caution the
reader than many other variants are possible and might in fact be
preferable.

\begin{definition}[$(n, k, W)$ Ensemble]
The $(n, k, W)$ ensemble is an ensemble of codes of rate $k/n$.
Each code is defined on $n$ streams and has $n-k$ shift-invariant
parity-checks. Each of these $n-k$ parity-checks has degree $n$
and it has exactly one tap for each of the $n$ streams. Each such
tap is picked independently of all other choices uniformly at random
from the set $\{0, \cdots, W-1\}$.  
\end{definition}

Let us go back to our previous definition of convolutional
codes. The parameter $W$ is typically close but always slightly
larger than the memory $M$ of the code.  Further, we see that each
code in this ensemble is defined by a parity-check matrix $\HP(D)$
whose entries $\h_{i, j}(D)$ are {\em monomials} where the degree
of the monomial is a random variable uniformly distributed in $\{0,
\cdots, W-1\}$. Figure~\ref{fig:diagram} below shows the standard
filter diagram where the number of input streams $n = 4$, the number
of shift-invariant parity-checks $k= 2$, and $W = 6$. Note that for
the ensembles we consider $W$ should be hundreds or thousands. The
$n$ and $k$ are typically small and only constrained by two facts:
First, the rate is equal to $k/n$. Second, as we will discuss in
the next section, the size of the minimum stopping set grows
(exponentially) in $n-k$ and so $n-k$ should not be too small.

\input{diagram.tex}

The reader might wonder why we are making this choice
and only allow monomials, whereas our previous generic definition
allowed polynomials as long as only a small number of taps is
non-zero. This choice is mainly motivated by the fact that this
ensemble is particularly easy to analyze when we are looking at the
size of the minimum stopping set. In practice even better performance
can likely be achieved by allowing several non-zero taps and we leave
the question of how to find ``optimal'' filter choices as an
interesting open problem.

As we mentioned before, the main difference to the standard definition
of LDPC convolutional codes in the literature is that the codes we consider are {\em
time-invariant}.  They are therefore defined by only a handful of
integer numbers. More precisely, each code in the ensemble $(n, k,
W)$ is determined by $ n (n-k)$ integer numbers in $\{0, \cdots,
W-1\}$. Hence, $n (n-k) \log_2 W$ bits sufficient to describe such
a code. Consequently, an exhaustive search for for the ``best" code in this ensemble 
needs to go over at most $W^{n(n-k)}$  codes. 
If we think of $W$ as the ``effective'' blocklength then
the construction complexity of such codes is polynomial in the
blocklength, namely, of order $W^{n (n-k)}$.

Encoding for members of the $(n, k, W)$ ensemble is particularly
simple if the code has the so-called ``stair-case'' property (defined
below). Recall that we have $n-k$ shift-invariant parity-checks, each
having $n$ taps, one on each stream. We can use the last $k$ streams
to contain information bits, and the bits on the first $n-k$ streams
can be evaluated deterministically using the bits on
the last $k$ streams. Assume now that we can order (label) the
shift-invariant parity-checks from $1$ to $n-k$ in a way that the associated
tap of the $i$-th check on the $i$-th stream is placed before the
tap of the $(i+1)$-th check on the $(i+1)$-th stream (we name this
configuration ``staircase''). The encoding can then be done in the
order determined by the ``staircase''.

%% file: diagram.tex
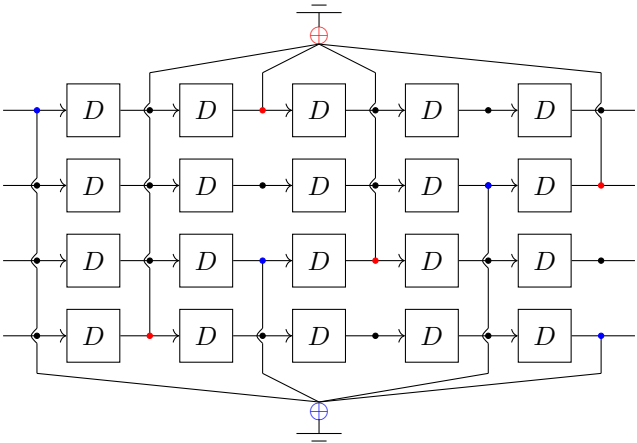
\begin{figure}[!htb]
\centering
\tikzset{
block/.style    = {draw, rectangle, minimum height = 2em, minimum width = 2em},
input/.style    = {coordinate},
output/.style   = {coordinate}}
\begin{tikzpicture}
%------------------------------------------------------------------------%

\node at (0,4)  [block] (strm4d1) {$D$};
\node at (1.5,4)[block] (strm4d2) {$D$};
\node at (3,4)  [block] (strm4d3) {$D$};
\node at (4.5,4)[block] (strm4d4) {$D$};
\node at (6,4)  [block] (strm4d5) {$D$};

\draw[->](-1.2,4)  -- (strm4d1);
\draw[->](strm4d1) -- (strm4d2);
\draw[->](strm4d2) -- (strm4d3);
\draw[->](strm4d3) -- (strm4d4);
\draw[->](strm4d4) -- (strm4d5);
\draw    (strm4d5) -- (7.2,4);

\filldraw[color = black] (0.75,4) circle (1pt);
\filldraw[color = black] (3.75,4) circle (1pt);
\filldraw[color = black] (5.25,4) circle (1pt);
\filldraw[color = black] (6.75,4) circle (1pt);

%------------------------------------------------------------------------%

\node at (0,3)  [block] (strm3d1) {$D$};
\node at (1.5,3)[block] (strm3d2) {$D$};
\node at (3,3)  [block] (strm3d3) {$D$};
\node at (4.5,3)[block] (strm3d4) {$D$};
\node at (6,3)  [block] (strm3d5) {$D$};

\draw[->](-1.2,3)  -- (strm3d1);
\draw[->](strm3d1) -- (strm3d2);
\draw[->](strm3d2) -- (strm3d3);
\draw[->](strm3d3) -- (strm3d4);
\draw[->](strm3d4) -- (strm3d5);
\draw    (strm3d5) -- (7.2,3);

\filldraw[color = black] (-0.75,3) circle (1pt);
\filldraw[color = black] (0.75,3) circle (1pt);
\filldraw[color = black] (2.25,3) circle (1pt);
\filldraw[color = black] (3.75,3) circle (1pt);
\filldraw[color = black] (5.25,3) circle (1pt);

%------------------------------------------------------------------------%

\node at (0,2)  [block] (strm2d1) {$D$};
\node at (1.5,2)[block] (strm2d2) {$D$};
\node at (3,2)  [block] (strm2d3) {$D$};
\node at (4.5,2)[block] (strm2d4) {$D$};
\node at (6,2)  [block] (strm2d5) {$D$};

\draw[->](-1.2,2)  -- (strm2d1);
\draw[->](strm2d1) -- (strm2d2);
\draw[->](strm2d2) -- (strm2d3);
\draw[->](strm2d3) -- (strm2d4);
\draw[->](strm2d4) -- (strm2d5);
\draw    (strm2d5) -- (7.2,2);

\filldraw[color = black] (-0.75,2) circle (1pt);
\filldraw[color = black] (0.75,2) circle (1pt);
\filldraw[color = black] (5.25,2) circle (1pt);
\filldraw[color = black] (6.75,2) circle (1pt);

%------------------------------------------------------------------------%

\node at (0,1)  [block] (strm1d1) {$D$};
\node at (1.5,1)[block] (strm1d2) {$D$};
\node at (3,1)  [block] (strm1d3) {$D$};
\node at (4.5,1)[block] (strm1d4) {$D$};
\node at (6,1)  [block] (strm1d5) {$D$};

\draw[->](-1.2,1)  -- (strm1d1);
\draw[->](strm1d1) -- (strm1d2);
\draw[->](strm1d2) -- (strm1d3);
\draw[->](strm1d3) -- (strm1d4);
\draw[->](strm1d4) -- (strm1d5);
\draw    (strm1d5) -- (7.2,1);

\filldraw[color = black] (-0.75,1) circle (1pt);
\filldraw[color = black] (2.25,1) circle (1pt);
\filldraw[color = black] (3.75,1) circle (1pt);
\filldraw[color = black] (5.25,1) circle (1pt);

%------------------------------------------------------------------------%

\node[color = red] at (3,5) (check1) {$\oplus$};
\draw (3,5.11) -- (3,5.3);
\draw (2.7,5.3) -- (3.3,5.3);
\draw (2.9,5.4) -- (3.1,5.4);

\draw (3,4.881) -- (0.75,4.5);
\draw (0.75,4.5) -- (0.75,4.1);
\draw (0.75,4.1) .. controls (0.65,4) .. (0.75,3.9);
\draw (0.75, 3.9) -- (0.75,3.1);
\draw (0.75,3.1) .. controls (0.65,3) .. (0.75,2.9);
\draw (0.75,2.9) -- (0.75,2.1);
\draw (0.75,2.1) .. controls (0.65,2) .. (0.75,1.9);
\draw (0.75,1.9) -- (0.75,1);
\filldraw[color = red] (0.75,1) circle (1pt);

\draw (3,4.881) -- (2.25,4.5);
\draw (2.25,4.5) -- (2.25,4);
\filldraw[color = red] (2.25,4) circle (1pt);

\draw (3,4.881) -- (3.75,4.5);
\draw (3.75,4.5) -- (3.75,4.1);
\draw (3.75,4.1) .. controls (3.65,4) .. (3.75,3.9);
\draw (3.75, 3.9) -- (3.75,3.1);
\draw (3.75,3.1) .. controls (3.65,3) .. (3.75,2.9);
\draw (3.75,2.9) -- (3.75,2);
\filldraw[color = red] (3.75,2) circle (1pt);

\draw (3,4.881) -- (6.75,4.5);
\draw (6.75,4.5) -- (6.75,4.1);
\draw (6.75,4.1) .. controls (6.65,4) .. (6.75,3.9);
\draw (6.75,3.9) -- (6.75,3);
\filldraw[color = red] (6.75,3) circle (1pt);

%------------------------------------------------------------------------%

\node[color = blue] at (3,0) (check2) {$\oplus$};
\draw (3,-0.11) -- (3,-0.3);
\draw (2.7,-0.3) -- (3.3,-0.3);
\draw (2.9,-0.4) -- (3.1,-0.4);

\draw (3,0.1190) -- (-0.75,0.5);
\draw (-0.75,0.5) -- (-0.75,0.9);
\draw (-0.75,0.9) .. controls (-0.85,1) .. (-0.75,1.1);
\draw (-0.75,1.1) -- (-0.75,1.9);
\draw (-0.75,1.9) .. controls (-0.85,2) .. (-0.75,2.1);
\draw (-0.75,2.1) -- (-0.75,2.9);
\draw (-0.75,2.9) .. controls (-0.85,3) .. (-0.75,3.1);
\draw (-0.75,3.1) -- (-0.75,4);
\filldraw[color = blue] (-0.75,4) circle (1pt);

\draw (3,0.1190) -- (2.25,0.5);
\draw (2.25,0.5) -- (2.25,0.9);
\draw (2.25,0.9) .. controls (2.15,1) .. (2.25,1.1);
\draw (2.25,1.1) -- (2.25,2);
\filldraw[color = blue] (2.25,2) circle (1pt);

\draw (3,0.1190) -- (5.25,0.5);
\draw (5.25,0.5) -- (5.25,0.9);
\draw (5.25,0.9) .. controls (5.15,1) .. (5.25,1.1);
\draw (5.25,1.1) -- (5.25,1.9);
\draw (5.25,1.9) .. controls (5.15,2) .. (5.25,2.1);
\draw (5.25,2.1) -- (5.25,3);
\filldraw[color = blue] (5.25,3) circle (1pt);

\draw (3,0.1190) -- (6.75,0.5);
\draw (6.75,0.5) -- (6.75,1);
\filldraw[color = blue] (6.75,1) circle (1pt);

%------------------------------------------------------------------------%
\end{tikzpicture}
\caption{Filter diagram: $n = 4$, $k = 2$, and $W = 6$.}
\label{fig:diagram}
\end{figure}

%% file: stopping_sets.tex
Recall that every code in the $(n, k, W)$ ensemble is defined by
only $n (n-k) \log_2(W)$ bits which determine the $n-k$ ``filters''.
The parity-checks are then defined by all shifts of these filters.
This means that there is a lot of ``dependence'' among the various
parity-checks. Does such a code have a large error floor?

We will investigate this question for the BEC by giving a bound on
the size of minimum stopping set, see \cite{DPRTU01}, for a typical
realization, where ``typical'' refers to the randomness in picking
the non-zero taps.

\begin{lemma}[Minimum Stopping Set Size]
\label{lem:minss} 
With probability $1 - O(1/W)$, every stopping set contained in
a randomly chosen code generated from the $(n,k,W)$ ensemble has
size at least $\bigl \lceil 3^{n-k} / \sqrt{2(n-k)}  \bigr \rceil$.
\end{lemma}
Note: The term $O(1/W)$ contains constants that depend on $n$. In
general, as will be clear from the proof, the larger $n$ the larger
we will have to choose $W$.

\begin{proof}
We begin by describing some basic properties of the positions of
the taps when $W$ is large enough. We will use
Figure~\ref{fig:example_proof} as our running example. Recall that
a code in the ensemble is specified by $n$ streams as well as $n-
k$ shift-invariant parity-checks. We denote the $i$-th shift-invariant
parity-check by $c_i$. From now on we will use the two phrases ``check
type $c_i$'' and ``shift-invariant parity-check $c_i$'' interchangeably.
A check type $c_i$ is determined by $n$ taps, one for each stream.
Let us denote by $v_{i, j}$ the tap on stream $j$ that is
associated to check $c_i$. From now on, we assume that the $i$-th
stream is placed on the horizontal line $y = i$ in the two dimensional
plane. The taps on each stream are placed at integer positions such that
any two consecutive taps differ by a unit. Consider two distinct
taps $v_{i, j}$ and $v_{i, j'}$ that are connected to check $c_i$. We
denote by $v_{i, j \to j'}$ the vector whose starting point is
$v_{i, j}$ and whose endpoint is $v_{i, j'}$ (see
Figure~\ref{fig:example_proof}). It is a $2$-dimensional vector
with integer components. The first component contains the difference
of the stream indices. The second component contains the difference
of the time indices.

\begin{figure}[h]
%\centering
\begin{tikzpicture}
\draw [color = red] (-9,-1) -- (-2, -1);
\draw [color = red] (-9,-2.5) -- (-2, -2.5);
\draw [color = red] (-9,-4) -- (-2, -4);

\node[scale = .9] at (-1.2, -.95) {\textcolor{red}{stream 1}};
\node[scale = .9] at (-1.2, -2.45) {\textcolor{red}{stream 2}};
\node[scale = .9] at (-1.2, -3.95) {\textcolor{red}{stream 3}};
	
\fill [fill=black] (-9,0) rectangle (-8.8,.2);

\node[scale = 1] at (-9, .5) {$c_i$};

\draw [color = black] (-8.9,.1) -- (-2.5, -1);

\fill [fill=black] (-2.5,-1) circle (.1cm);

\node[scale = 1] at (-2.5 , -.6) {$v_{i,1}$};

\draw [color = black] (-8.9,.1) -- (-7, -2.5);

\fill [fill=black] (-7, -2.5) circle (.1cm);

\node[scale = 1] at (-7 , -2.1) {$v_{i,2}$};

\draw [color = black] (-8.9,.1) -- (-8, -4);

\fill [fill=black] (-8, -4) circle (.1cm);

\node[scale = 1] at (-7.8 , -3.6) {$v_{i,3}$};

\draw [color = blue, -{>[scale = 2.5, length = 2, width = 2]}] (-2.5, -1) -- (-6.92, -2.46);
\draw [color = blue, -{>[scale = 2.5, length = 2, width = 2]}] (-2.5, -1) -- (-7.92, -3.96);

\node[scale = 1] at (-4.9 , -1.5) {$v_{i,1 \to 2}$};
\node[scale = 1] at (-6 , -3.4) {$v_{i,1 \to 3}$};

%\fill [fill=black] (-10,-5) rectangle (-9.8,-4.8);

%\draw [color = black] (-9.9,-4.9) -- (-3, -1);
%\draw [color = black] (-9.9,-4.9) -- (-5, -2.5);
%\draw [color = black] (-9.9,-4.9) -- (-9, -4);
\end{tikzpicture}
\caption{\label{fig:example_proof}}
\end{figure}

It is not hard to show that, with probability $1 - O(1/W)$, all the
vectors $v_{i, j \to j'}$ are distinct. Consider now a stopping set
and assume without loss of generality (w.l.o.g.) that it contains a variable
node on the first stream. Let us denote this variable by $z$. Recall
now that each check has exactly one connection to each stream.
Hence, for each $i \in \{1,2, \cdots, c=n-k\}$, there is a check
node of type $i$ (appropriately shifted) that is connected to $z$.
But whenever this happens, the corresponding check must have at
least one more non-zero variable that it is connected to at this
time, since otherwise we do not have a stopping set. Equivalently,
we can say that for each $i$, there exists a vector $v_{i, 1 \to
j}$ such that if we start at $z$, and move along $v_{i, 1 \to j}$,
then we end up at a variable node which is also part of the stopping
set. This itself already leads to a lower bound on the size of a
stopping set (namely the bound $1+c=1+n-k$) since, as we mentioned
above, the various vectors are with high probability distinct. 

But we can get better bounds by continuing to ``grow out'' the stopping
set (think of a tree rooted at $z$). So assume that we start at the
variable $z$. This variable has $c$ children
that are distinct with high probability. Now let us look at the
children of these children and so on, up to depth $\ell$, $\ell
\leq c$.

More precisely, given any sequence of \textit{distinct} check types $(i_1,
i_2, \cdots, i_\ell)$, we can associate a sequence of vectors $(v_{i_1,
1 \to j_1}, v_{i_2, j_1 \to j_2}, \cdots, v_{i_\ell, j_\ell \to
j_{\ell+1}})$ such that if we start at node $z$ and move along the
path created by these vectors then all the nodes that we visit along
this path belong to the stopping set. If all these nodes were all
distinct (for all such paths of distinct types) we would get a very
simple lower bound on the stopping set. As we will see now, it can
happen that some of the nodes are in fact the same. But we will be
able to lower bound the number of distinct such nodes. 

Define the set $T$ as
\begin{equation} \nonumber
T = \big\{ (i_1, i_2, \cdots, i_\ell) \mid \ell \in [0,c], i_1, \cdots, i_\ell \text{ are distinct} \big\}.
\end{equation}
Note that we allow the empty (null) sequence to be included in $T$.
We now construct a rooted tree (rooted in whose vertices are members
of $T$. This tree has the empty string as its root node, and every
$ (i_1, i_2, \cdots, i_\ell) \in T$  is adjacent to $(i_1, i_2, \cdots,
i_{\ell-1})$ as its parent node. In this way the depth of a node $(i_1,
i_2, \cdots, i_\ell) \in T$ is equal to $\ell$. Consider any path $P$ in
the tree that starts at the root node and ends at a sequence $(i_1,
i_2, \cdots, i_\ell) \in T$. Define $\bar{i}_j \triangleq (i_1, i_2,
\cdots, i_j)$ for each $j \in [\ell]$. We also denote the root node
by $\bar{i}_0$. Therefore, the path $P$ can be represented as $P =
\bar{i}_0 - \bar{i}_1 - \cdots - \bar{i}_\ell$.

Recall from above that we can assume w.l.o.g.~that the stopping set
has a node on the first stream which we denote by $z$. From what
we have discussed up to now, given any stopping set we can assign
to each sequence $\bar{i} \in T$ a number
$n_{\bar{i}} \in [n]$ such that the following property holds. For
any path $P = \bar{i}_0 - \bar{i}_1-\cdots-\bar{i}_\ell$, consider the
following trajectory: we start at node $z$ and move along the vectors
$v_{i_{j}, n_{\bar{i}_{j-1}} \to n_{\bar{i}_j}}$ for each $j \in[\ell]$,
one after the other, then all the variable nodes that we visit along
the way should belong to the stopping set.

%As a result, for each sequence $\bar{i} \in T$ we can associate a
%variable $v_{\bar{i}}$ in the stopping set using the following
%procedure. Consider a path $P$ that starts from the root node and
%ends at a sequence $\bar{i}_{\ell} = (i_1, \cdots, i_\ell) \in T$. We associate
%to $P$ a path of variable nodes in the stopping set which starts
%from $z$ and walks along the directions $v_{i_{j}, n_{\bar{i}_{j-1}}
%\to n_{\bar{i}_j}}$ for $j \in [\ell]$, and the final variable that
%we reach (after $\ell$ translations) is $v_{\bar{i}_{\ell}}$.

We need to count the number of repetitions among the variables
$v_{\bar{i}}$ for $\bar{i} \in T$ (and this leads to a lower bound
on the size of the stopping set). Consider two sequences $\bar{i}
\triangleq (i_1, i_2, \cdots, i_\ell) \in T$ and $\bar{h} \triangleq
(h_1, h_2, \cdots, h_{\ell'}) \in T$. With probability $ 1 - O(1/W)$,
we have $v_{\bar{i}} = v_{\bar{h}}$ only if $\ell = \ell'$ and the two
sets of vectors $\{  v_{ {i_j}, n_{\bar{i}_{j-1}} \to n_{\bar{i}_j}
} \}_{j \in [\ell]}$ and $\{ v_{{h_j}, n_{\bar{h}_{j-1}} \to n_{\bar{h}_j
}  } \}_{j \in [\ell]}$ are the same (up to a permutation). With this
condition, for a sequence $\bar{i}$, the number of repetitions of
$v_{\bar{i}}$ can be upper bounded as follows. For $t \in
[n]$, let
\begin{equation*} 
x_t = \big| \{j \in [\ell]  :  n_{\bar{i}_j} = t\}  \big|.
\end{equation*}  
Note that $\sum_{t \in [n]} x_t = \ell$. The number of repetitions of $v_{\bar{i}}$ is upper bounded by $\prod_{t \in [n]} x_t!$. We can thus upper bound the number of repetitions by
\begin{equation*}
B_\ell \triangleq \max \bigg\{\prod_{t \in [n]} x_t! \, \,\mathlarger{\mid} \, \, (x_1, x_2, \cdots, x_n): \sum_{t \in [n]} x_t = \ell \bigg\}.
\end{equation*} 
By using the inequality $a! b! \leq (a+b)!$, we can further write $B_\ell \leq \max\{ a! b! \mid a+b = \ell  \}$, and thus
\begin{equation} 
\label{B_k_upper}
B_\ell \leq \left\{
\begin{array}{ll}
\big(\frac{\ell}{2}!\big)^2  & \text{if } \ell \text{ is even},\\
\\
\big(\frac{\ell-1}{2}!\big) \big(\frac{\ell+1}{2}!\big)& \text{if } \ell \text{ is odd}.
\end{array} \right.
\end{equation}
Finally, the number of sequences $\bar{i} \in T$ with length $\ell$
is precisely $c ! / (c-\ell)!$. Putting all these together, we
obtain the following lower bound on the size of any stopping set
(which holds with probability $1 - O(1/W)$):
\begin{align*}
\sum_{\ell \in [c]} \frac{1}{B_\ell}\frac{c!}{(c - \ell)!} &\geq \sum_{\ell \in [c]} \frac{2^\ell}{\sqrt{2\ell}} { {c}\choose{\ell}} \\
& \geq \frac{1}{\sqrt{2c}} \sum_{\ell \in [c]} 2^\ell {{c}\choose{\ell}}  \geq \frac{3^c}{\sqrt{2c}},
\end{align*}
where we have used \eqref{B_k_upper} and Stirling's bounds.
\end{proof}
%\begin{itemize}
%\item [(P1)] Assume that $e$ connects two sequences $(i_1, i_2, \cdots, i_{k-1})$ and $(i_1, i_2, \cdots, i_{k-1}, i_k)$. Then the associated vector to $e$ should be of the form $v_{i_k, j \to k}. $ 
%\item [ (P2)] Consider two edges $e_1, e_2 \in E$ that are both 
%\end{itemize}

%% file: simulation.tex
We now get to our numerical  experiments.  More precisely, we
consider the ``terminated'' case: on each stream the number of
variable nodes is $10^6$, and on both sides of each stream $W$
additional variable nodes are fixed to be $0$ (seeding).
%\footnote{In practice, when we are concerned about the rate loss, we would likely
%only use a one-sided termination. But our main concern here is to
%show that such simple codes still show the threshold saturation
%phenomenon and so the exact termination scheme that we use is secondary.} 
We assume that the all-zero codeword was transmitted
over the binary erasure channel (BEC) with parameter $\epsilon$ and
we use the peeling decoder, see \cite{RiU08}.

\begin{figure}[!htb]
\centering
\includegraphics[width=6cm]{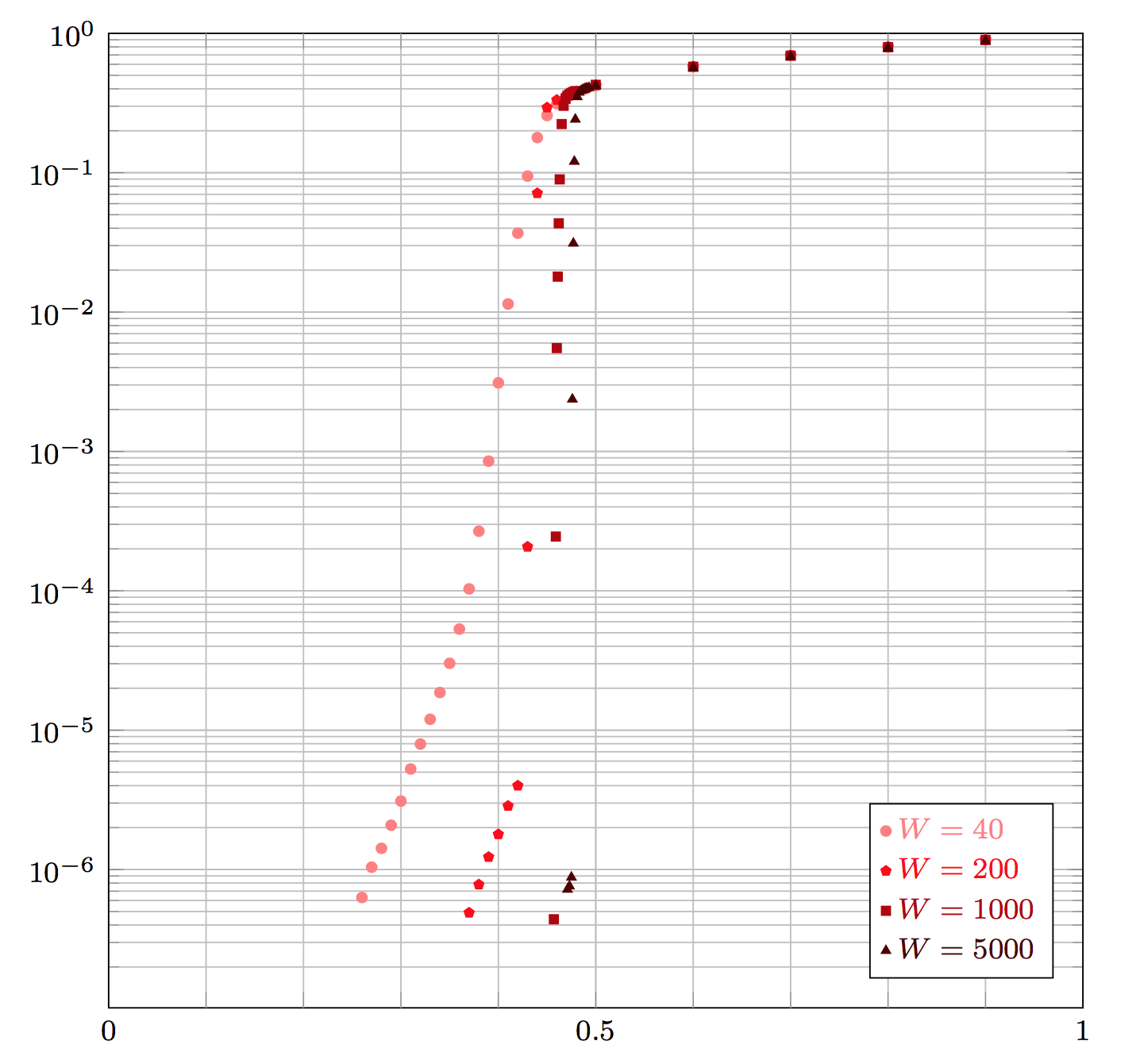}
\caption{$n = 6$ and $W \in \{40, 200, 1000, 5000\}$.}
\label{fig:n6}
\end{figure}

Figure~\ref{fig:n6} shows the empirical bit erasure probability
versus the channel parameter $\epsilon$, for the cases $n = 6$ and
$W \in \{40, 200, 1000, 5000\}$. The empirical average is over $100$
random samples from the ensemble (where the randomness is over the
choice of code and the channel realization). As we can see, the
curves become steeper and steeper in the waterfall region and move
rightward as $W$ becomes larger and larger. This is consistent with
the fact that $W$ is proportional to the ``effective'' block length.

%The error floor region tends to vanish as $W$ increases. 
%More concretely, for $W = 5000$, the empirical
%probability of bit erasure in the normal scale is $7.20000 \times
%10^{-7}$ when $\epsilon = 0.475$, it goes to $2.36803 \times 10^{-3}$
%when $\epsilon = 0.476$, it is $2.41446 \times 10^{-1}$ when $\epsilon
%= 0.479$, and it is $3.99265 \times 10^{-1}$ when $\epsilon = 0.489$.
%The figure reveals a threshold saturation phenomenon and as we can
%see in the large system size limit the iterative decoding threshold
%of the ensemble tends to the MAP threshold of the $(3,6)$-regular
%LDPC ensemble.

In addition we see in this figure the error floor due to small
stopping sets. We can compare this error floor to our analytic
predictions. Note that the ``slope'' of the error floor curve tells
us the size of the stopping set that causes this error floor. I.e.,
if the curve has the form $\alpha \epsilon^d$ then $d$ is the size
of the stopping set. For $W = 40$ we get an estimate of $d \approx 12$
and for $W = 1000$ we have the estimate $d \approx 15$. These results
are consistent with our lower bound of Lemma \ref{lem:minss} in
Section \ref{sec:ss}, since $\lceil 3^3 / \sqrt{6} \rceil = 12$.

\begin{figure}[!htb]
\centering
\includegraphics[width=6cm]{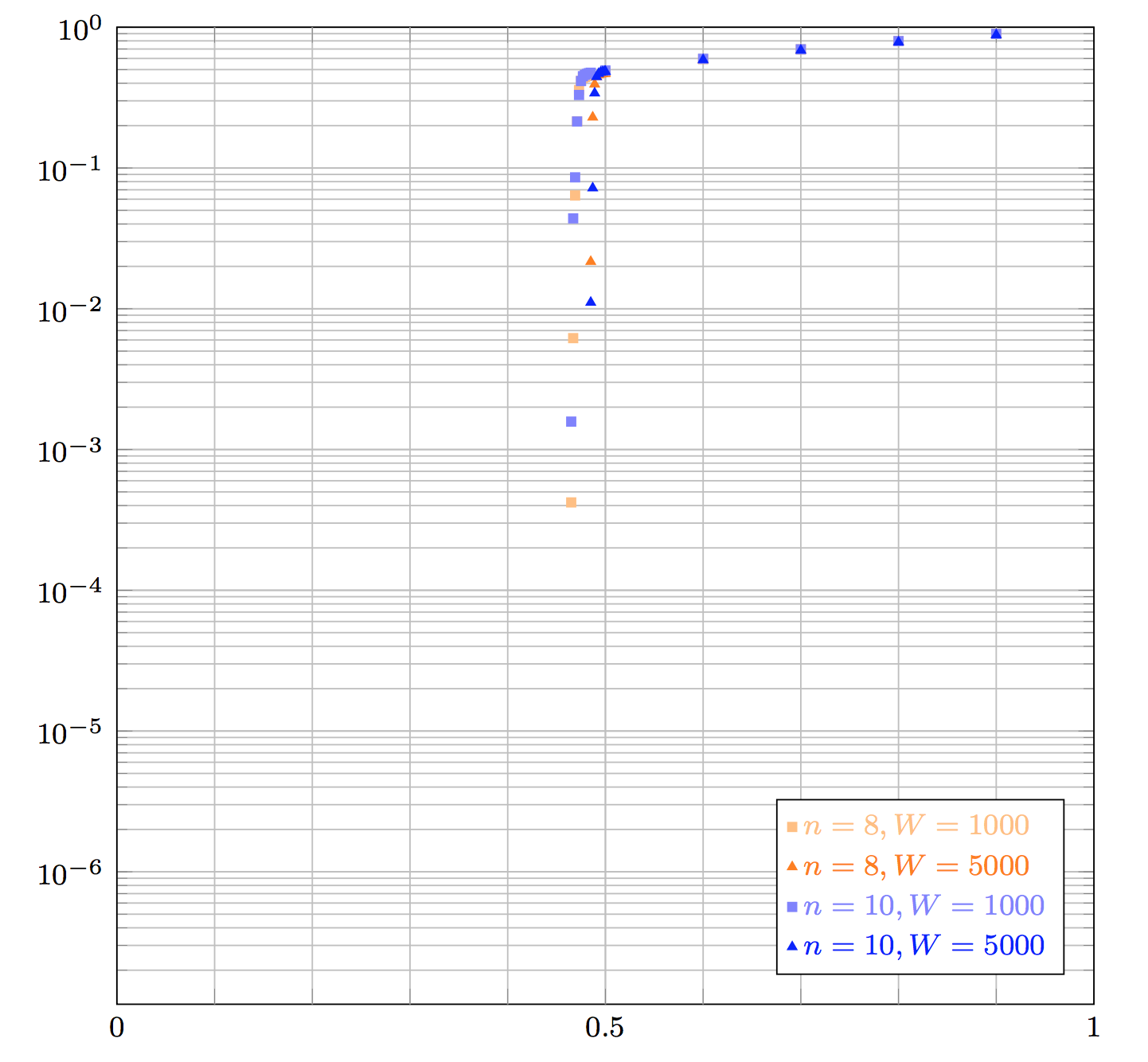}
\caption{$n \in \{8, 10\}$ and $W \in \{1000, 5000\}$.} 
\label{fig:n8and10}
\end{figure}

The cases $n \in \{8, 10\}$ and $W \in \{1000, 5000\}$ are shown
in Figure~ \ref{fig:n8and10}. Note that in order to see the error
floor in this case one would have to simulate the curves to
considerably lower probabilities since even for $n=8$ and $k=4$ we
have already $\lceil 3^4 / \sqrt{8} \rceil = 29$. We see that compared to the case $n = 6$ the threshold is even closer
to the Shannon capacity, consistent with the threshold saturation
phenomenon.

% the error floors are
%hardly observed in these cases.  Take $n = 10$ and $W = 5000$ as
%an example.  The empirical probability of erasure is equal to $0$
%when $\epsilon$ is less than or equal to $0.485$; it is $6.14285
%\times 10^{-2}$ when $\epsilon = 0.487$. Again, as one can expect,
%for fixed $n$, the iterative decoding threshold of our ensemble
%tends to the MAP threshold of the corresponding regular LDPC ensemble
%as $W$ tends to $\infty$. Increasing $n$ in the large $W$ limit
%finally allows the sequence of thresholds approaches the Shannon
%capacity.

